# Modeling of graphene Hall effect sensors for microbead detection

A. Manzin[1], E. Simonetto[1,2], G. Amato[1], V. Panchal[3] and O. Kazakova[3]

[1]Istituto Nazionale di Ricerca Metrologica (INRIM), Torino, Italy

[2]Dipartimento di Elettronica e Telecomunicazioni, Politecnico di Torino, Torino, Italy

[3]National Physical Laboratory (NPL), Teddington, UK

Corresponding author:

Dr. Alessandra Manzin

Istituto Nazionale di Ricerca Metrologica

Strada delle Cacce, 91

I-10135 TORINO

ITALY

Tel. +39-011-3919825

Email: a.manzin@inrim.it

# ABSTRACT

This paper deals with the modeling of sensitivity of epitaxial graphene Hall bars, from sub-micrometer to micrometer size, to the stray field generated by a magnetic microbead. To demonstrate experiment feasibility, the model is first validated by comparison to measurement results, considering an ac-dc detection scheme. Then, an exhaustive numerical analysis is performed to investigate signal detriment caused by material defects, saturation of bead magnetization at high fields, increment of bead distance from sensor surface and device width increase.

## I. INTRODUCTION

In the last decade, many studies have been focused on the fabrication and characterization of miniaturized semiconductor Hall sensors for biomedical applications, as the detection of magnetic nanoparticles and microbeads for biological labeling and drug delivery.[1] Recently, it has been shown that graphene can be an advantageous alternative to semiconductor materials for the development of high-sensitivity devices, due to its high room temperature carrier mobility, low-cost production and possibility of reducing the vertical distance between active layer and target.[2-7] Different device performances are expected, depending on the Hall bar geometrical properties and the graphene fabrication process,[2] e.g. chemical vapor deposition, molecular beam epitaxy on SiC and exfoliation.

In this work, we explore the possibility of using miniaturized graphene devices as detectors of magnetic beads with micrometer size. To this aim, we simulate the graphene sheet as a 2D electronic system in the diffusive regime, by means of a finite element model that provides the spatial distribution of the electric potential inside the Hall plate in presence of a strongly localized magnetic field (i.e. the bead stray field).[8]

The model is first validated by comparison to experimental results obtained on an epitaxial graphene sensor (with intrinsic n-type doping) mimicking an ac-dc detection scheme.[5] Then, it is applied to investigate the possibility of using graphene Hall bars for bead susceptibility mapping,[9] by varying device width, dc field and bead vertical distance. Finally, we study the degradation of device performance due to material defects and heterogeneities in the electrical properties, e.g. bi-layer islands[10] and topographic corrugations[11,12], modeling graphene as a multi-carrier system.[13]

## II. NUMERICAL MODEL

To include non-homogeneous conductivity properties due to multi-layer islands and topological defects, the graphene sample is modeled as a 2D electronic system with $M$ types of carriers with distinct mobility and density.[13] Under the assumptions of diffusive transport regime and non-uniform orthogonal magnetic field, $\mathbf{B} = B(x,y)\mathbf{k}$, charge transport is described by means of the Ohm's law $\mathbf{J} = \ddot{\sigma}\mathbf{E}$, having introduced a spatially dependent conductivity tensor $\ddot{\sigma}$, whose elements are

$$\begin{cases} \sigma_{xx} = \sigma_{yy} = \sum_{i}^{M} \frac{\Gamma_i n_i e \mu_i}{1+\mu_i^2 B^2} \\ \sigma_{xy} = -\sigma_{yx} = \sum_{i}^{M} \frac{\Gamma_i n_i e \mu_i^2 B}{1+\mu_i^2 B^2} \end{cases} \quad (1)$$

where $e$ is the electron charge and $\Gamma_i$ is the characteristic function associated to the $i$-th carrier type with mobility $\mu_i$ and density $n_i$. By expressing the electric field $\mathbf{E}$ as a function of scalar potential $\phi$ ($\mathbf{E} = -\nabla\phi$) and considering the equation of continuity for current density vector $\mathbf{J}$, it follows that

$$\nabla \cdot \left[ \ddot{\sigma}(x,y) \nabla \phi(x,y) \right] = 0 \quad (2)$$

Problem (2), which is completed by ad-hoc boundary conditions at the current and voltage contacts and at the insulating boundaries, is solved by applying the finite element method.

To simulate ac-dc Hall magnetometry techniques,[5,9] the magnetic field in (1) incorporates a uniform external field, perpendicular to the sensor surface and composed of a dc and an ac signals ($\mathbf{B}_{ext} = \mathbf{B}_{dc} + \mathbf{B}_{ac}$). Moreover, it includes the orthogonal component of the stray field produced by a magnetic bead assumed to be uniformly magnetized along the direction of $\mathbf{B}_{ext}$. The bead, represented as a magnetic dipole, is responsible for the generation of a field with $z$-component

$$B_{bead_z}(x,y) = \frac{\mu_0}{4\pi} m(B_{ext}) \left[ \frac{3d^2}{r^5} - \frac{1}{r^3} \right] \quad (3)$$

where $r$ is the distance between the point of calculus in the graphene plane, with coordinates $(x,y)$, and the barycentre of the bead with moment amplitude $m$ and distance $d$ above the graphene device.

The ac-dc detection scheme is mimicked by calculating the amplitude of the ac Hall voltage due to the bead ($V_{ac,bead}$) for a specific dc field. The ac contribution is extrapolated by approximating $m$ as

$$m(B_{ext}) \cong m(B_{dc}) + B_{ac} \frac{V}{\mu_0} \chi(B_{dc}) \tag{4}$$

where $\chi$ is the magnetic susceptibility of the bead.

### III. MODEL VALIDATION

The numerical model is validated by comparison to experimental results obtained at room temperature on a 1600 nm wide Hall bar made of epitaxial graphene grown on 4H-SiC(0001) and with three-cross configuration (see inset of Fig. 1). The material, with intrinsic n-type doping, has carrier density $n = 1.07 \times 10^{16}$ m$^{-2}$ and mobility $\mu = 0.17$ m$^2$/Vs.[5] To remove parasitic inductive signals, the device sensitivity is experimentally characterized via an ac-dc Hall magnetometry technique, based on the measurement of the in-phase component of the ac Hall voltage.[5] An iron oxide bead with 1 μm diameter and mass saturation magnetization of ~22 Am$^2$/Kg (Dynal, MyOne)[14,15] is placed on top of one cross, and its detection is performed by applying a step dc field varying between $B_{dc}^0 = 0$ T (duration of 60 s) and $B_{dc}^1 = 0.25$ T (duration of 30 s), fixing $B_{ac}$ to 3 mT. The bead presence leads to a change in the amplitude of the in-phase component of the ac signal ($\Delta V_{ac,bead}$), which is proportional to the variation in bead magnetic susceptibility.

Figure 1 reports the measured step amplitudes for the cross with bead and the empty one, together with a comparison with the calculated signal, where $\chi(B_{dc}^0)$ and $\chi(B_{dc}^1)$ are directly derived from the magnetization curve in Ref. [14] to avoid susceptibility underestimation at 0 T correlated to the use of Langevin function.[9] A good agreement between experimental and simulation results is found, with a variation in $V_{ac,bead}$ higher than 6 μV (the calculated value of $V_{ac,bead}$ at 0 T is 6.5 μV).

## IV. NUMERICAL ANALYSIS

A parametric analysis of bead susceptibility mapping[9] is carried out considering epitaxial Hall bars constituted by two symmetric crosses with variable width $w$; the longitudinal and transversal arms have a length equal to $18w$ and $9w$, respectively (inset of Fig. 2). The sensor performances are studied by varying the dc magnetic field and the bead distance from graphene sheet, also investigating the impact of possible material heterogeneities, as presence of islands of bi-layer graphene and topographic corrugations (substrate terraces). For all cases, $B_{ac}$ is fixed to 10 mT and the frequency to 400 Hz, considering constant-current-supply or current mode as well as constant-voltage-supply or voltage mode.[3]

*A. Influence of device width and bead distance*

The analysis starts from Hall devices with cross width $w$ ranging from 400 nm to 1500 nm and made of a monolayer epitaxial graphene. The considered material has homogeneous carrier density $n = 2 \times 10^{16}$ m$^{-2}$ and room-temperature mobility $\mu = 0.3$ m$^2$/Vs,[4,16] corresponding to an electron mean free path of ~70 nm that justifies the hypothesis of diffusive transport regime. The analysis is performed in current mode, setting the bias current at 75 µA. For all the device sizes, the 4-terminal resistance and the Hall coefficient are ~20 kΩ and 312 Ω/T, respectively. In the high-frequency thermal noise range, the voltage-noise spectral density is ~18 nV/$\sqrt{\text{Hz}}$, which corresponds to a minimal detectable field of ~0.8 µT/$\sqrt{\text{Hz}}$ at 75 µA. However, in the usual working frequency range, $1/f$ or flicker noise becomes the dominant contribution, limiting sensitivity performance when reducing the device size and increasing the bias current.[5] If we assume a Hooge parameter of ~$10^{-4}$,[17,18] an operating frequency of 400 Hz and $I_{bias} = 75$ µA, the spectral density rises up to ~3 µV/$\sqrt{\text{Hz}}$ for the 400 nm width device and to ~0.8 µV/$\sqrt{\text{Hz}}$ when $w = 1500$ nm.

The magnetic moment resolution is strongly affected by the dc magnetic field $B_{dc}$ applied to magnetize the bead. This is well demonstrated by Fig. 2, which shows the role of $B_{dc}$ on the amplitude of the ac Hall voltage ($V_{ac,bead}$) due to a bead placed above the device cross centre at a vertical distance $d$ of 500 nm between its barycentre and the graphene sheet, for different values of cross width $w$. The non-linearity of the bead magnetization curve, which saturates at ~0.8 T, leads to a rapid diminution in the device response with the dc field increase, following the bead susceptibility decay. A reduction of a factor of ~22 is observed for all the range of variation of $w$ when varying $B_{dc}$ from 0.1 T to 0.5 T, approaching white-noise level for larger devices. For the entire considered interval of $B_{dc}$, the sensitivity reduces of a factor of ~9 when increasing the device width from 400 nm to 1500 nm. To avoid signal degradation caused by non-linearity and to operate above flicker noise, low dc fields (~0.1 T) should be considered.

Figure 3 analyzes the effect of the increase in bead vertical distance $d$ for different cross widths $w$. Strong signal detriment is found for $w = 400$ nm, with a reduction to one third when the bead is moved from the sensor surface to a position $d = 800$ nm, in correspondence of which flicker noise is reached. Conversely, a slower variation is obtained for larger devices, guaranteeing to operate above flicker noise in a wide distance interval.

*B. Influence of material defects*

The presence of multi-layer islands can affect the electronic transport properties and thus the device performance. As an example, we study the sensitivity of a Hall bar with $w = 1000$ nm, containing a bi-layer region with hexagonal shape that is responsible for a local increase in carrier concentration (inset of Fig. 4a). The material properties of mono-layer graphene are the same of the previous analysis, while the bi-layer island is assumed to have the same mobility of the bulk and a carrier density 8 times higher.[10] The role of the material defect position is shown in Fig. 4a, by comparing current and voltage modes and placing the bead at the cross centre in contact with the device surface; the size of the bi-layer region is ~575 nm (diameter of the circumscribed circle) and $B_{dc} = 0.1$ T.

In current mode ($I_{bias}$ = 75 µA), which is strongly dependent on carrier density,[3] there is a significant decay (~30%) in the signal amplitude when the defect is located at the cross centre, while its effect becomes negligible for positions outside the cross junction area. For the centered position, $V_{ac,bead}$ varies from 3.3 µV to 1.5 µV when increasing the defect size from 145 nm to 1150 nm. Voltage mode, which is generally affected by the only carrier mobility when material uniformity is considered, displays here a quantitative behavior similar to the current mode, due to perturbation in the current density spatial distribution induced by material heterogeneity (Fig. 4b). Epitaxial graphene grown on SiC can be also characterized by topographic corrugations due to the surface morphology of the substrate, which results in periodic structures consisting of terraces and step edges.[11,12] These heterogeneities, caused by preparation processes, lead to anisotropic electron transport that can strongly affect the sensitivity to localized magnetic fields. The role of periodic inhomogeneities is here investigated by considering a 1000 nm wide Hall bar with terraces and steps aligned both perpendicular (Fig. 5a) and parallel (Fig. 5b) to the device channel. The local material properties are derived from the experimental results reported in Ref. [11] at zero gate voltage, i.e. terraces have carrier density $n_T$ of $7.5 \times 10^{16}$ m$^{-2}$, mobility $\mu_T$ of 0.1 m$^2$/Vs and variable width $w_T$ up to 160 nm, while steps have carrier density $n_S$ of $9.4 \times 10^{16}$ m$^{-2}$, mobility $\mu_S$ of 0.01 m$^2$/Vs and fixed width $w_S$ of 10 nm. The estimated spectral density associated to 1/$f$ noise when $I_{bias}$ = 75 µA and $f$ = 400 Hz is ~0.47 µV/$\sqrt{Hz}$ for a device with homogeneous material properties equal to the terrace ones.

Figure 5d compares device sensitivity for the two considered material microstructures as a function of terrace width, analyzing both types of supply. In current mode, there is a weak dependence on $w_T$ and microstructure, as a consequence of the small change in carrier density between terrace and step regions. A non-negligible signal reduction is found for $w_T$ lower than 60 nm. Conversely, a strong influence of the step-terrace orientation is observed for the voltage mode, as a result of the charge transport anisotropy induced by the high variation in carrier mobility. The perpendicular orientation leads to a more rapid decay of device performance when reducing $w_T$, reaching the estimated flicker noise when $w_T \sim 50$ nm.

The obtained results can be interpreted by modelling charge transport along the parallel (perpendicular) direction via an equivalent electrical circuit consisting of terrace and step resistances connected in parallel (series).[11] According to this approximation, the effective electrical properties along the two orientations can be expressed as $a_{eff//} = \gamma a_T + (1-\gamma) a_S$ and $\frac{1}{a_{eff\perp}} = \frac{\gamma}{a_T} + \frac{(1-\gamma)}{a_S}$, where γ is the fraction of terrace region and $a$ stands for $n$ or μ. The strong influence of material microstructure on device sensitivity found for voltage mode is strictly dependent on the deviation of the effective mobilities along the two directions for low values of γ (Fig. 5c).

## V. CONCLUSIONS

A numerical model has been developed, validated and applied to study the sensitivity of graphene devices in presence of localized stray fields generated by magnetic microbeads, simulating ac-dc detection scheme. Larger signals have been generally found in comparison to semiconductor devices based on InSb due the possibility of using high bias currents. However, the analysis has demonstrated how the performances of graphene devices can deteriorate in presence of material defects and how only a proper selection of experimental parameters (sensor width, supply mode and dc field for bead magnetization) can guarantee high detectable signals.

**ACKNOLEDGEMENTS**

This work has received funding from the Italian MIUR Project "Metrology for therapeutic and diagnostic techniques based on electromagnetic radiation and ultrasound waves" (2014-2016).

**FIGURE CAPTIONS**

FIG. 1. Measured and calculated variation in the amplitude of the in-phase component of the ac Hall voltage in response to a step dc field (amplitude of 250 mT and duration of 30 s), considering a 3 mT ac field at 210 Hz. The inset shows the scheme of the considered Hall bar with width $w$ = 1600 nm.

FIG. 2. Amplitude of the ac Hall voltage due to a bead placed in contact with the sensor surface at the cross centre, as a function of $B_{dc}$ and device width $w$ ($I_{bias}$ = 75 µA). The inset shows the scheme of the considered Hall bars.

FIG. 3. Amplitude of the ac Hall voltage as a function of bead vertical distance $d$, for different device width $w$ ($I_{bias}$ = 75 µA). The dc field is set at 0.1 T.

FIG. 4. (a) Amplitude of the ac Hall voltage due to a bead placed in contact with the sensor surface at the cross centre, as a function of position $u$ of a bi-layer island with 575 nm size (inset). (b) Spatial distribution of current density vector perturbed by defect presence.

FIG. 5. Schemes of the Hall bars with terraces and steps aligned both perpendicular (a) and parallel (b) to the device channel. (c) Effective mobility and carrier density as a function of terrace width $w_T$. (d) Amplitude of the ac Hall voltage due to a bead placed in contact with the sensor surface at the cross centre versus $w_T$, for both material microstructures and $B_{dc}$ = 0.1 T. The voltage value imposed in voltage mode is the one associated to a current of 75 µA, when the material has homogeneous properties equal to terrace ones.

**FIGURES**

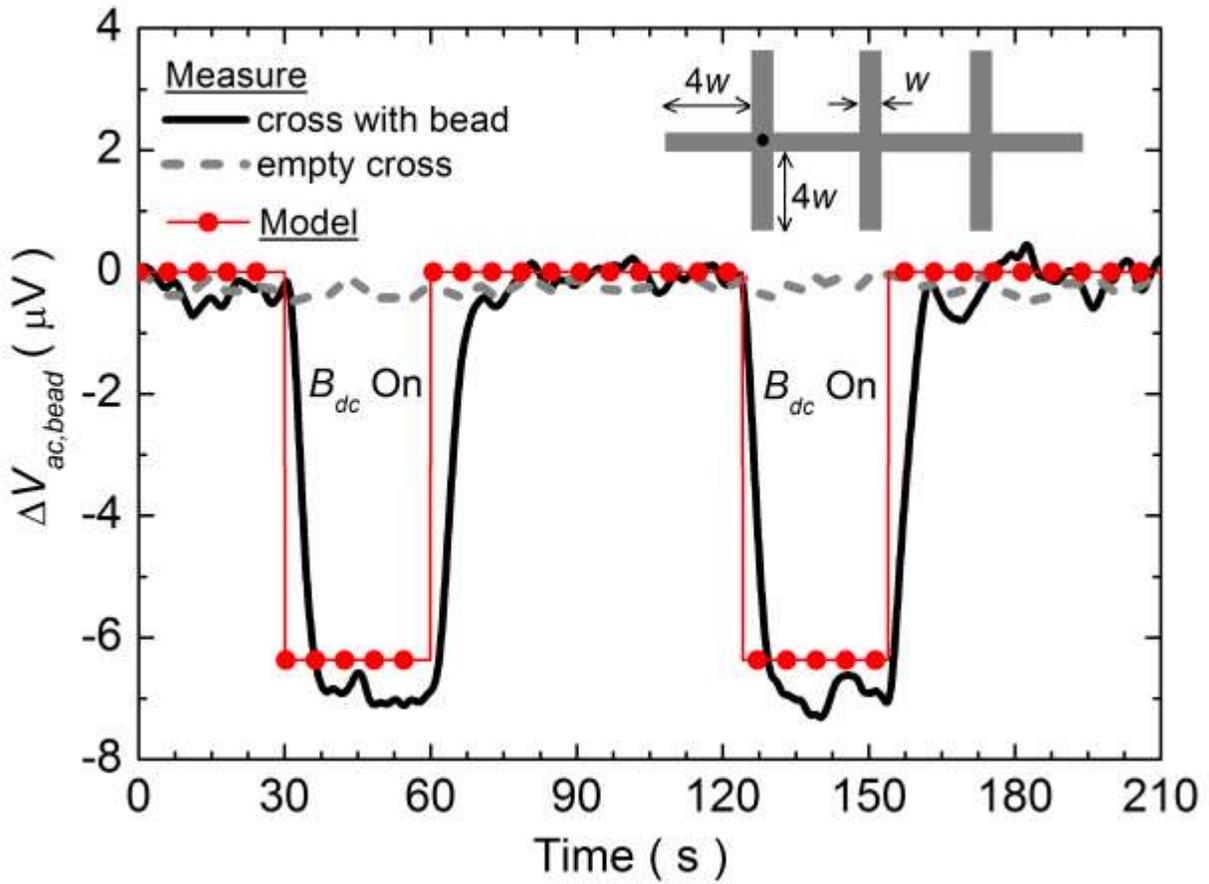

FIG. 1. Measured and calculated variation in the amplitude of the in-phase component of the ac Hall voltage in response to a step dc field (amplitude of 250 mT and duration of 30 s), considering a 3 mT ac field at 210 Hz. The inset shows the scheme of the considered Hall bar with width $w$ = 1600 nm.

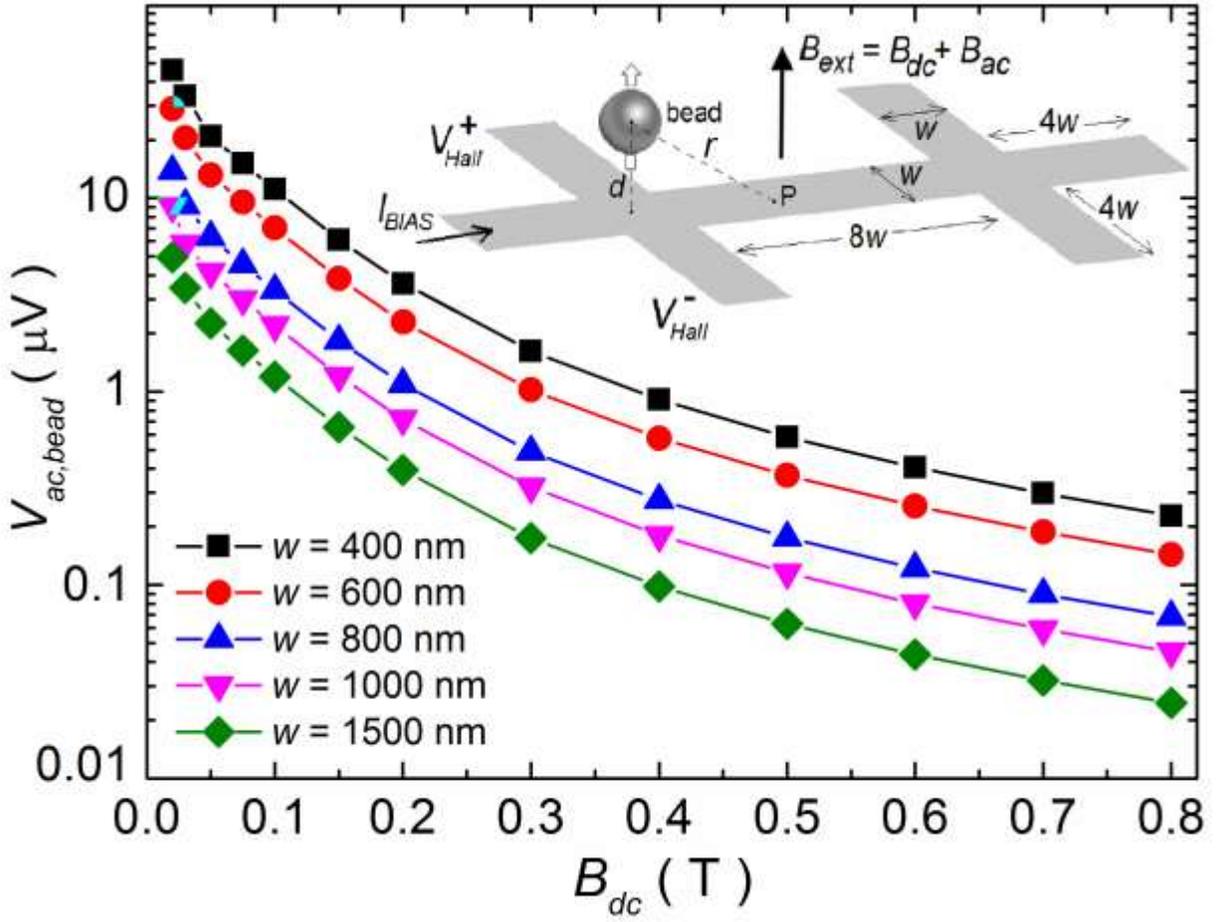

FIG. 2. Amplitude of the ac Hall voltage due to a bead placed in contact with the sensor surface at the cross centre, as a function of $B_{dc}$ and device width $w$ ($I_{bias}$ = 75 μA). The inset shows the scheme of the considered Hall bars.

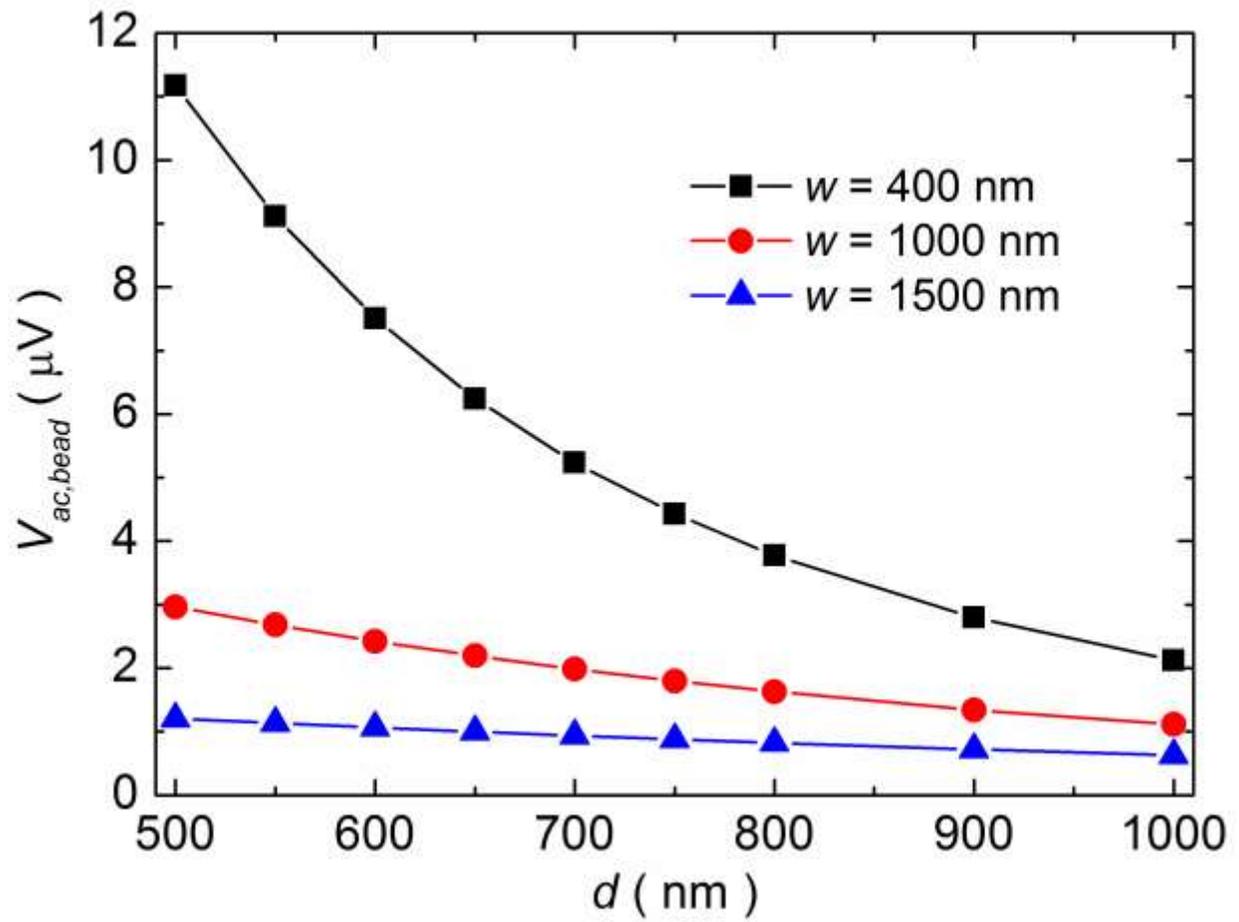

FIG. 3. Amplitude of the ac Hall voltage as a function of bead vertical distance *d*, for different device width *w* ($I_{bias}$ = 75 μA). The dc field is set at 0.1 T.

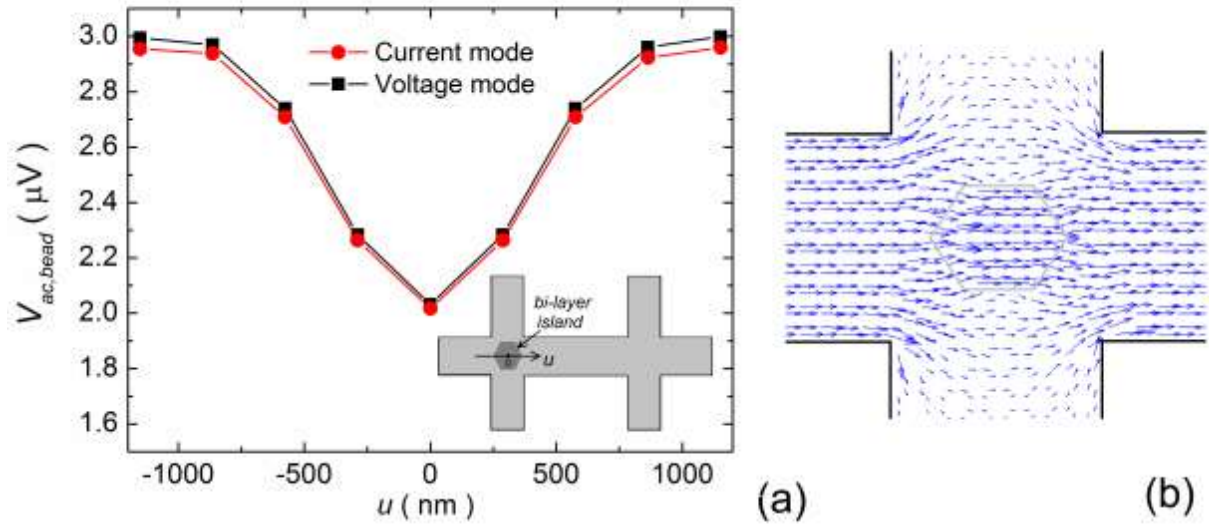

FIG. 4. (a) Amplitude of the ac Hall voltage due to a bead placed in contact with the sensor surface at the cross centre, as a function of position $u$ of a bi-layer island with 575 nm size (inset). (b) Spatial distribution of current density vector perturbed by defect presence.

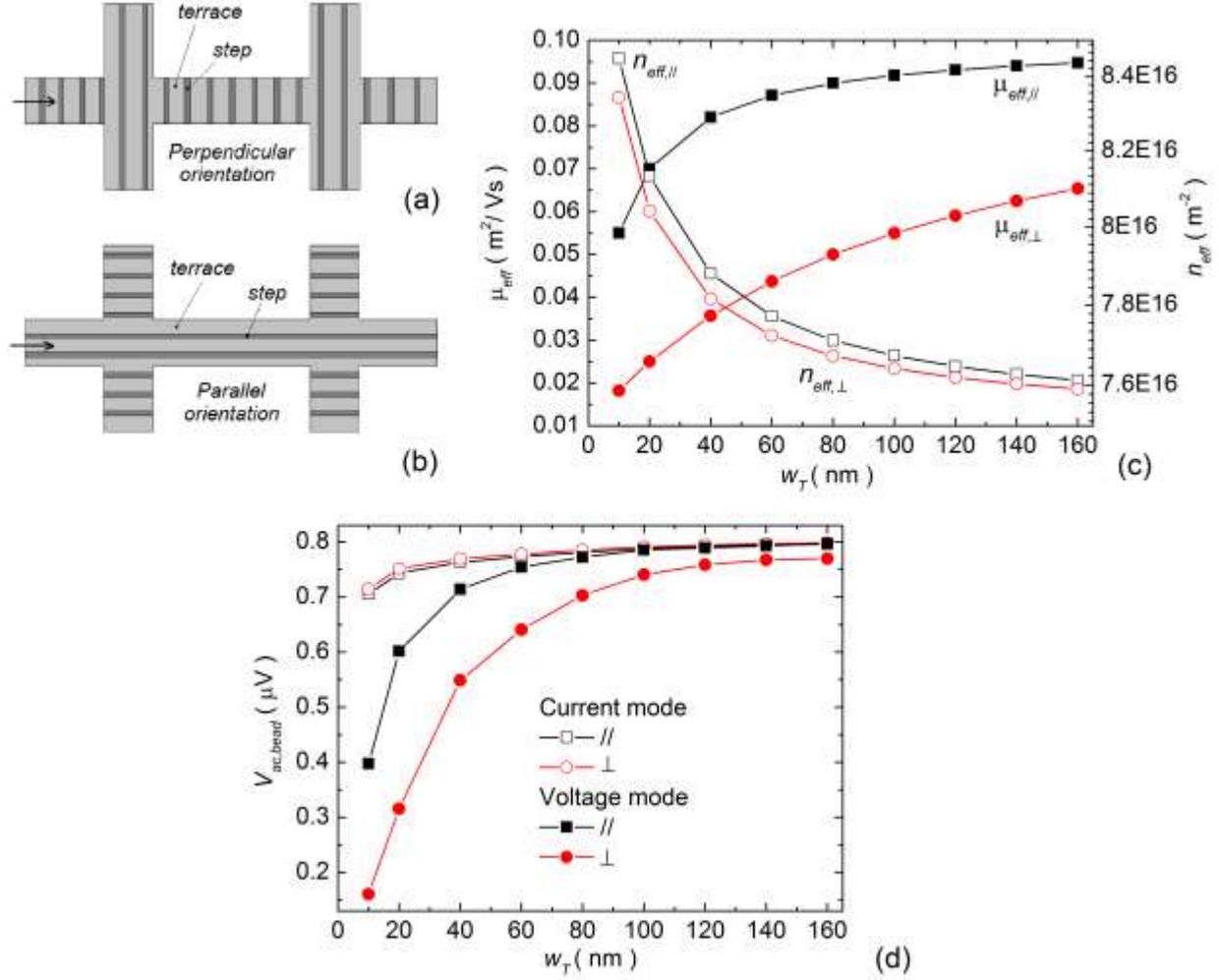

FIG. 5. Schemes of the Hall bars with terraces and steps aligned both perpendicular (a) and parallel (b) to the device channel. (c) Effective mobility and carrier density as a function of terrace width $w_T$. (d) Amplitude of the ac Hall voltage due to a bead placed in contact with the sensor surface at the cross centre versus $w_T$, for both material microstructures and $B_{dc} = 0.1$ T. The voltage value imposed in voltage mode is the one associated to a current of 75 μA, when the material has homogeneous properties equal to terrace ones.